# Hierarchical Cellular Structures in High-Capacity Cellular Communication Systems

Prof. R. K. Jain, Sumit Katiyar
Research Scholar, Singhania University,
Jhunjhunu, Rajasthan, India

Prof. N. K. Agrawal Sr M IEEE
Inderprastha Engineering College, Sahibabad
Ghaziabad, India

*Abstract*—In the prevailing cellular environment, it is important to provide the resources for the fluctuating traffic demand exactly in the place and at the time where and when they are needed. In this paper, we explored the ability of hierarchical cellular structures with inter layer reuse to increase the capacity of mobile communication network by applying total frequency hopping (T-FH) and adaptive frequency allocation (AFA) as a strategy to reuse the macro and micro cell resources without frequency planning in indoor pico cells [11]. The practical aspects for designing macro- micro cellular overlays in the existing big urban areas are also explained [4]. Femto cells are inducted in macro / micro / pico cells hierarchical structure to achieve the required QoS cost effectively.

*Keywords- Hierarchical cellular structures; Total frequency hopping; Adaptive frequency allocation; Communication system traffic; Code division multiple access (CDMA).*

## I. INTRODUCTION

The provision of capacity for the increasing traffic demand in mobile radio networks comes along with the reduction of the cell size and, hence, stronger traffic fluctuations between the cells. Moreover, improved indoor coverage is required. Hierarchical cellular structures can serve indoor users and hot spots by pico- and micro cell layers, respectively, while providing coverage in the area by the macro cell layer. Moreover, hierarchical cellular structures can compensate traffic fluctuations e.g. by shifting overflow traffic from lower to higher layers. In order to avoid interference between the layers, their frequency allocations have to be coordinated. This can be achieved by incorporating smart antenna / Intelligent antenna in hierarchical structure with adaptive-SDMA approach.

Moreover, hierarchical cellular structures become a regular feature of future mobile radio networks. Although different multiple access techniques may apply, some experiences from GSM can also be useful for the design of other hierarchical cellular networks, where several layers share the same resources [1]. The results of [1], comparative simulation study has been discussed, which aims at network configurations in a dense urban environment where a high additional traffic capacity in the pico cell layer shall be achieved solely by reusing the micro- and/or macro cell frequencies. The results of number of general conclusions for the design of hierarchical cellular networks are also discussed [1].

In order to obtain a useful knowledge about the deployment and operation of macro- micro CDMA cellular overlays, we must deal with the existing conditions of today's big urban areas, which include spatial and temporal traffic distributions, geographical characteristics, user mobility characteristics, and so on. In [5] a novel algorithmic approach for the joint deployment of macro cells and micro cells over big urban areas having spatially non-uniform traffic distributions has been discussed. Based on a discrete area representation, the proposed algorithm determines the locations, radii and required capacities of macro cells and micro cells, which guarantee the required quality of service (QoS) cost effectively. For the practical design of macro- micro CDMA cellular overlays we have to take care of the following issues in depth-

i) Effect of user's mobility.

ii) In presence of many high mobility users, the undesirable micro cellular coverage holes in hot spot areas incur a critical problem, assuming a significant volume of limited macro- cellular code division multiple access (CDMA) capacity.

iii) Even though a macro- micro cellular overlay well accommodates the traffic loads of working hours, it may loose much of its significance if it can not manage a temporal traffic variation during a day.

iv) Since only a few wideband CDMA carriers exist in the third generation wireless personal communications, it is quite difficult to fully exploit all the potentials of macro- micro cellular overlays.

v) If some operating functions, such as cell layer selections or interlayer handovers, are improperly configured, they can degrade the performance of a macro-micro cellular overlay considerably.

Since the number of mobile users is continuously growing, we shrink cell size to increase system capacity. By shrinking cell size, handoff rate is increased. To overcome these problems, hierarchical cell structure is proposed. As efficient use of radio resources is very important, utilization of all resources have to be optimized. Thus, in hierarchical cell structure, how to assign available radio resources to each user is critical question [6]. In order to adapt the changes of traffic, adaptive radio resource management can be considered in CDMA based hierarchical cell structure. The proposed scheme improves call blocking, call dropping and optimal utilization of radio resources.

## II.   TECHNOLOGIES FOR ENHANCEMENT OF SPECTRAL DENSITY IN HIERARCHICAL SYSTEM

A tractable, flexible and accurate model for downlink heterogeneous cellular networks (fig 1) was developed successfully. The model consist of K tiers of randomly located base stations where each tier may differ in terms of average transmit power, the supported data rate and BS density. This allows elements spanning traditional, micro, pico, and femtocell BSs to be simultaneously considered. Assuming a mobile user connects to its strongest BS, we derive its Signal-to-Interference-Ratio (SIR) distribution and use to find the coverage (equivalently outage) probability over the entire network. The accuracy of these analytical results through empirical comparisons with an actual 4G macro cell network verified [7]. Cellular networks are becoming increasingly heterogeneous due to the co-deployment of many disparate infrastructure elements, including micro, pico and femtocells, and distributed antennas. A flexible, accurate and tractable model for a general downlink HCN consisting of K tiers of randomly located BSs, where each tier may differ in terms of average transmit power, supported data rate, and BS density. Assuming 1) a mobile connects to the strongest BS, 2) the target Signal to- Interference-Ratio (SIR) is greater than 0 dB, and 3) received power is subject to Rayleigh fading and path loss. Expressions for the average rate achievable by different mobile users are derived. This model reinforces the usefulness of random spatial models in the analysis and research of cellular networks. This is a baseline tractable HCN model with possible future extensions being the inclusion of antenna sectoring, frequency reuse, power control and interference avoidance/ cancellation [8].

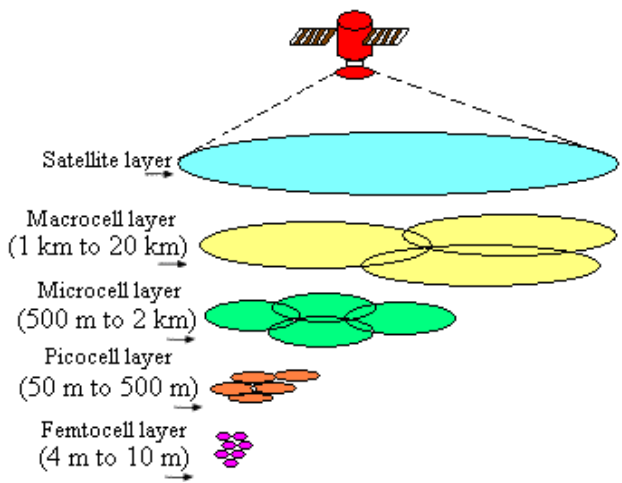

Satellite layer

Macrocell layer
(1 km to 20 km)

Microcell layer
(500 m to 2 km)

Picocell layer
(50 m to 500 m)

Femtocell layer
(4 m to 10 m)

Figure 1.   Hierarchical Cellular Structure

To overcome handoff problem in hierarchical cell structure, efficient use of radio resources is very important. All resources have to be optimally utilized. However, in order to adapt to changes of traffic, it is necessary to consider adaptive radio resource management. An adaptive radio resource management in CDMA based hierarchical cell structure was proposed. In this scheme, the resource shortage in micro cell is solved by increasing the number of resource and the resource shortage in macro cell is solved by decreasing traffic in macro cell. One aspect of this proposed scheme is to increase threshold velocity rather than to decrease threshold velocity when most channels in micro cell are busy. Thus, abrupt increase of macro cell load caused by decreasing threshold velocity when micro cell is overloaded can be solved. Though the number of handoff is increased in this case, the number of call loss will decrease. Since this problem occurs only during rush hours. It will prevent call from dropping or blocking although handoff rate is increased slightly. The result of proposed scheme demonstrates improvement in call dropping, call blocking and utilization of resource [6].

1) Practical designs of macro-micro CDMA cellular overlays in the existing big urban areas (having especially non-uniform traffic distributions) were proposed. The numerical results by extensive event-driven simulations show that the resulting macro-micro cellular overlays successfully cope with the existing conditions of today's big urban areas, such as spatial and temporal traffic distributions and user mobility characteristics [9].

2) The ability of hierarchical cellular structures with inter-layer reuse to increase the capacity of a GSM (Global System for Mobile Communications) radio network by applying Total Frequency Hopping (T-FH) and Adaptive Frequency Allocation (AFA) as a strategy to reuse the macro- and micro cell resources without frequency planning in indoor picocells have been discussed. The presented interference analysis indicates a considerable interference reduction gain by T-FH in conjunction with AFA, which can be used for carrying an additional indoor traffic of more than 300 Erlang/km2, i.e. increasing the spectral efficiency by over 50 %, namely 33 Erlang/km2/MHz. From these results, it can be concluded that hierarchical structures required reuse strategies that not only adapt to the current local interference situation, but additionally distribute the remaining interference to as many resources as possible. For a hierarchical GSM network this requirement is fulfilled by the T-FH/AFA technique very well [11].

3) The design of a cellular network is a complex process that encompasses the selection and configuration of cell sites and the supporting network infrastructure. This investigation presents a net revenue maximizing model that can assist network designers in the design and configuration of a cellular system. The integer programming model takes as given a set of candidate cell locations with corresponding costs, the amount of available bandwidth, the maximum demand for service in each geographical area and the revenue potential in each customer area. Based on these data, the model determines the size and location of cells, and the specific channels to be allocated to each cell. To solve problem instances, a maximal clique cut procedure is developed in order to efficiently generate tight upper bounds. A lower bound is constructed by solving the discrete optimization model with some of the discrete variables fixed. Computational experiments on seventy-two problem instances demonstrate the computational viability of our new procedure [10].

4) This paper proposes a combined channel assignment (CCA) mechanism for hierarchical cellular systems with overlaying macro cells and overlaid micro cells. The proposed CCA mechanism combines overflow, underflow, and reversible schemes, where new or handoff calls having no available channel to use in the overlaid micro cell can overflow to use free channels in the overlaying macro cell, handoff calls from a neighboring macro cell can underflow to use free channels in the overlaid micro cell, and handoff attempts from a macro cell only region to an overlaid micro cell can be reversed to use free channels in the micro cell. We apply the CCA mechanism in two different hierarchical cellular systems of Strip type and Manhattan type and compare the CCA with the overflow channel assignment (OCA) scheme. Simulation results show that the CCA mechanism outperforms the OCA scheme by once in forced termination probability, by several times in new call blocking probability, and by 4.7% in system utilization for a hierarchical cellular system, and the CCA mechanism is more suitable for the Manhattan type than for the Strip type [1].

5) In this paper, we develop a vision for the future of wireless communications beyond the third generation, which consists of a combination of several optimized access systems on a common IP-based medium access and core network platform. Different access systems inter-work via horizontal (intra-system) and vertical (inter-system) handover, service negotiation and global roaming. These complementary access systems are optimized for different applications and environments. They are allocated to different cell layers in the sense of hierarchical cells with respect to cell size, coverage and mobility to provide globally optimized seamless services for all users. New air interfaces can also be incorporated to satisfy demands for higher data rates, increased mobility and reduced cost per bit. This vision requires international research and standardization activities to solve many technical challenges. Key issues include the global interworking of different access systems on a common platform, advanced antenna concepts and the implementation of multi-mode and multi-band terminals as well as base stations via software-defined radio concepts [12].

6) The surest way to increase the system capacity of a wireless link is by getting the transmitter and receiver closer to each other, which creates the dual benefits of higher quality links and more spatial reuse. In a network with nomadic users, this inevitably involves deploying more infrastructures, typically in the form of micro cells, hotspots, distributed antennas, or relays. A less expensive alternative is the recent concept of femtocells—also called home base-stations—which are data access points installed by home users to get better indoor voice and data coverage. In this article, author overviewed the technical and business arguments for femtocells, and described the state-of-the-art on each front. Authors also described the technical challenges facing femtocell networks, and gave some preliminary ideas for how to overcome them [13].

The demand for cellular radio services growing rapidly, and in heavy populated areas the need arises to shrink the cell sizes and scale the coverage pattern. The extension of the service into the PCN domain, railway stations, malls, pedestrian areas, markets and other hotspots further enhances this trend. The vision of future cellular systems incorporates macro, micro, pico and femto cells in hierarchical structure (fig 2).

Architectures proposed introduce the concept of remoting the antennas from the cell site, thus allowing for many micro cells to be served by a single attended center. One scheme advocates a linear transformation of the RF to optical frequencies and relaying the signals via fibers to a center. Others propose a down conversion to IF (70 MHz) which is then relayed by microwaves or by fibers.

### A. Macro Cell

A conventional base station with 20W power and range is about 1 km to 20 km. Macro cell in hierarchical structure takes care of roaming mobiles.

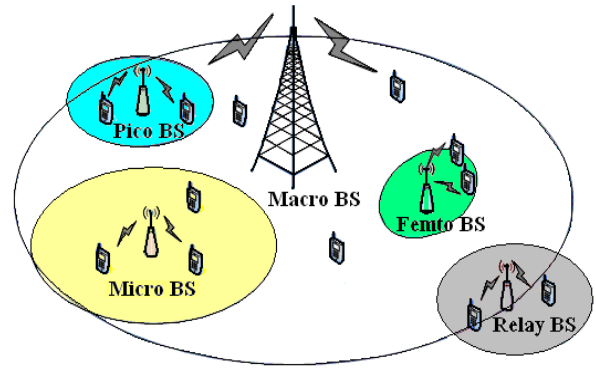

Figure 2. An example of single RAT hierarchical (Multi-tier) architecture framework

### B. Micro Cell

A conventional base station with 1W to 5W power and range is about 500 m to 2 km. Micro cells and pico cells takes care of slow traffic (pedestrian and in-building subscribers). Micro cells can be classified as the following:

1) Hot Spots: These are service areas with a higher tele-traffic density or areas that are poorly covered. A hot spot is typically isolated and embedded in a cluster of larger cells.

2) Downtown Clustered Micro cells: These occur in a dense, contiguous area that serves pedestrians and mobiles. They are typically found in an "urban maze of" street canyons," with antennas located far below building height.

3) In-Building. 3-D Cells: These serve office buildings and pedestrians (fig 3). This environment is highly clutter dominated, with an extremely high density and relatively slow user motion and a strong concern for the power consumption of the portable units.

### C. Pico Cell

The picocells are small versions of base stations, ranging in size from a laptop computer to a suitcase.

Besides plugging coverage holes, picocells are frequently used to add voice and data capacity, something that repeater and distributed antenna cannot do.

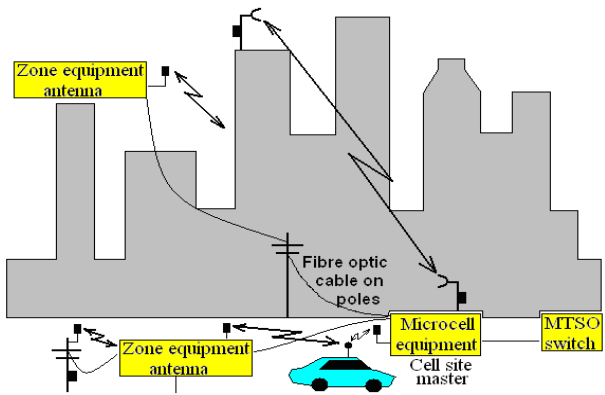

Figure 3.    Microcell Installation Concept

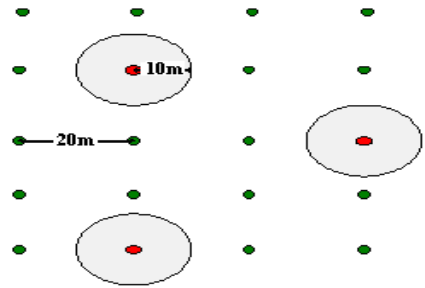

Figure 4.    Illustration of FAP deployment

Adding capacity in dense area, splitting cells are expensive, time consuming and occasionally impossible in dense urban environment where room for a full size base station often is expensive or unviable. Compact size picocells makes them a good fit for the places needing enhanced capacity, they can get.

Picocells are designed to serve very small area such as part of a building, a street corner, malls, railway station etc. These are used to extend coverage to indoor area where outdoor signals do not reach well or to add network capacity in areas with very dense uses.

### D.  Femto Cell

A femtocell is a smaller base station, typically designed for use in home or small business. In telecommunications, a femtocell is a small cellular base station, typically designed for use in a home or small business. It connects to the service provider's network via broadband (such as DSL or cable); current designs typically support 2 to 4 active mobile phones in a residential setting, and 8 to 16 active mobile phones in enterprise settings. A femtocell allows service providers to extend service coverage indoors, especially where access would otherwise be limited or unavailable (fig 4). Although much attention is focused on WCDMA, the concept is applicable to all standards, including GSM, CDMA2000, TD-SCDMA, Wi-MAX and LTE solutions. Femtocells are also called home base stations which are data access points installed by home users to get better indoor voice and data coverage. Data networks require much higher signal quality in order to provide the multi-Mbps data rates to individual user in the prevailing scenario (fig 5) and this can be met by using femtocells in hierarchical structure (macro, micro, pico cells along with femto cell for home user). This hierarchical structure will also result in better coverage, improved capacity, improved macro / micro cell reliability, reduced cost, reduced power consumption and RF pollution.

The capacity benefits of femtocells are:

1. Reduced distance between the femtocell and the user, which leads to a higher received signal strength. This will result in improvements in capacity through increase signal strength and reduced interference.

2. Lowered transmit power, and mitigation of interference from neighboring macro cell and femto cell users due to outdoor propagation and penetration losses. This will result in improvements in capacity through increased signal strength and reduced interference.

3. As femtocells serve only around 1-4 users, they can devote a larger portion of their resources (transmit power & bandwidth) to each subscriber. A macro / micro cell, on the other hand, has a larger coverage area (500m-20 km radius), and a larger number of users; providing Quality of Service (QoS) for data users is more difficult. Deploying femtocell will enable more efficient uses of precious power and frequency resources.

### III.    Smart / Adaptive Antenna

The adoption of smart antenna techniques in future wireless systems is expected to have a significant impact on the efficient use of the spectrum, the minimization of the cost of establishing new wireless networks, the optimization of service quality, and realization of transparent operation across multi technology wireless networks. Smart antennas have emerged as potentially a leading technology for achieving highly efficient networks which maximize capacity and improve quality and coverage.

Smart antenna can provide greater capacity and performance benefits than standard antennas because they can be used to customize and fine-tune antenna coverage patterns that match the traffic conditions in a wireless network or that are better suited to complex radio frequency (RF) environments. Furthermore, smart antennas provide maximum flexibility by enabling wireless network operators to change antenna patterns to adjust to the changing traffic or RF conditions in the network [16].

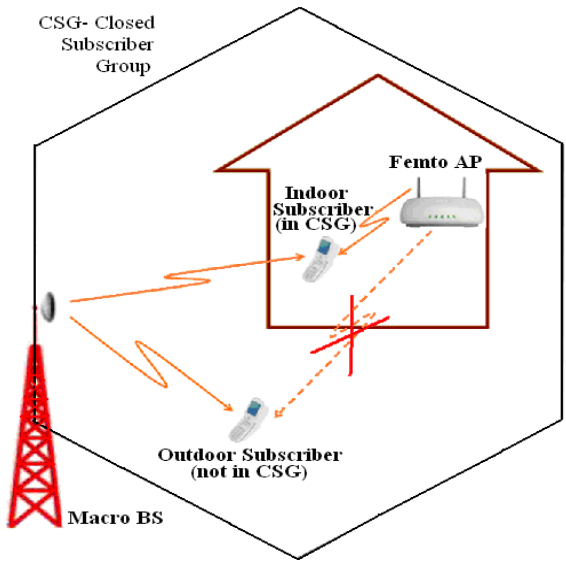

Figure 5. CSG Cell Association

Smart antennas at base stations can be used to enhance mobile communication systems in several ways:

- Increased BS range

- Less interference within the cell

- Less interference in neighboring cells

- Increased capacity by means of SFIR or SDMA

'Smart' antenna transmitters emit less interference by only sending RF power in the desired directions. Furthermore, 'smart' antenna receivers can reject interference by looking only in the direction of the desired source. Consequently 'smart' antennas are capable of decreasing CCI. A significantly reduced CCI can be taken as advantage of Spatial Division Multiple Access (SDMA) [9]. The same frequency band can be re-used in more cells, i.e. the so called frequency re-use distance can be decreased. This technique is called Channel Re-use via Spatial Separation. *In essence, the scheme can adapt the frequency allocations to where the most users are located [19].* With the inclusion of nanotechnology devices, the capability of adaptive antenna will increase manifold.

## IV. APPLICATION OF NANOTECHNOLOGY

Nano- technology could provide solutions for sensing, actuation, radio, embedded intelligence into the environment, power efficient computing, memory, energy sources, man-machine interaction, materials, mechanics, manufacturing & environmental issues.

Nanotechnology will rapidly boost all these disciplines and their application areas. Economic impact is foreseen to be comparable to information technology and telecom industries. One of the central visions of wireless industry aims at ambient intelligence: computation and communication always available and ready to serve the user in an intelligent way, so that we may optimally utilize the scarcest resource- frequency spectrum and achieve faster speeds (i.e., enhanced data rate).

Mobile devices together with the intelligence that will be embedded in human environments- home, office, public places- will create a new platform that enables ubiquitous sensing, computing, storage and communication. Core requirements for this kind of ubiquitous ambient intelligence are that the devices are autonomous and robust. They can be deployed easily and require little maintenance. As data rates require more memory and computing shown in fig 6, mobile devices will be the gateways to personally access ambient intelligence and needed information. Mobile also implies limited size and restrictions on the power consumption. Seamless connectivity with other devices and fixed networks is a crucial enabler for ambient intelligence system- this leads to requirements for increased power, which together with the size limitations leads to severe challenges in thermal management. All above requirements can be addressed satisfactorily with the application of nanotechnology.

Nanotechnology may augment the sensory skills of a human based on wearable or embedded sensors and the capabilities to aggregate this immense global sensory data into meaningful information for our everyday life.

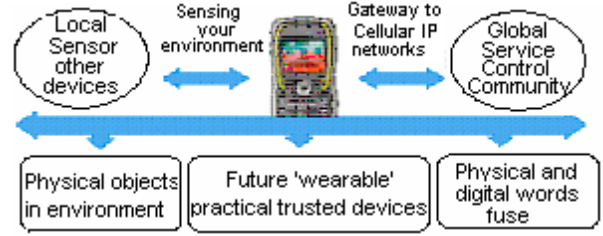

Figure 6. Mobile devices become gateway to ambient intelligence and needed information

Nanotechnology can help to develop novel kind of intelligent devices where learning is one of the key characteristic properties of the system, similarly to biological systems which grow and adapt to the environment autonomously [17].

## V. HANDOFF IN CELLULAR SYSTEM

Handoff is an essential element of cellular communications. Efficient handoff algorithms are a cost-effective way of enhancing the capacity and QoS of cellular systems. Macro cell radii are in several kilometers. Due to the low cell crossing rate, centralized handoff is possible despite the large numbers of MSs are to be managed by MSC. The use of micro cells is considered the single most effective means of increasing the capacity of cellular systems. Micro cells are more sensitive to the traffic and interference than macro cells due to short term variations (e.g., traffic and interference variations), medium / long term variations (e.g., new buildings), and incremental growth of the radio network (e.g., new BSs) [18]. The number of handoffs per cell is increased by an order of magnitude, and the time available to make a handoff is decreased [19]. Using an umbrella cell is one way to reduce the handoff rate. Due to the increase in the micro cell boundary crossings and expected high traffic loads, a higher degree of decentralization of the handoff process becomes necessary [20]. Lot of handoff schemes has been developed. The schemes packages show that these schemes can significantly decrease both the number of dropped handoff calls and the number of blocked calls without degrading the quality of communication service and the soft handoff process [18], [19], [20], [21], [22], [23], [24], [25], [26].

Handoff is an integral component of cellular communications. Efficient handoff algorithms can enhance system capacity and service quality cost effectively.

## VI. PROPOSED NETWORK

Proposed network (Fig 7 & 8) is based on simple handoff algorithm discussed in [25] and hierarchical cellular structures with inter-layer reuse in an enhance GSM radio network is suggested [1]. A design of Macro-Micro CDMA Cellular Overlays in the Existing Big Urban Areas is suggested [4]. It is also assumed that the MS is equipped with a Rake receiver capable of performing "maximal ratio combining" of the signals it receives from the transmitting BSs [26]. The following cellular structure will be used for dense urban areas [27]:

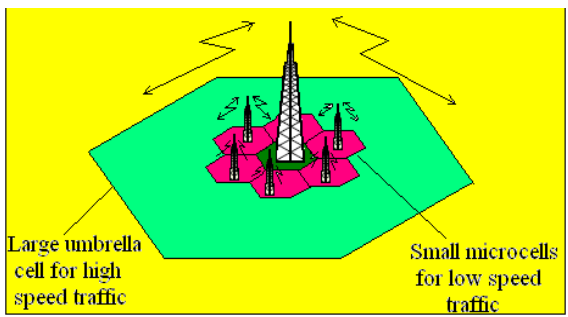

Figure 7.  The Umbrella Cell Approach

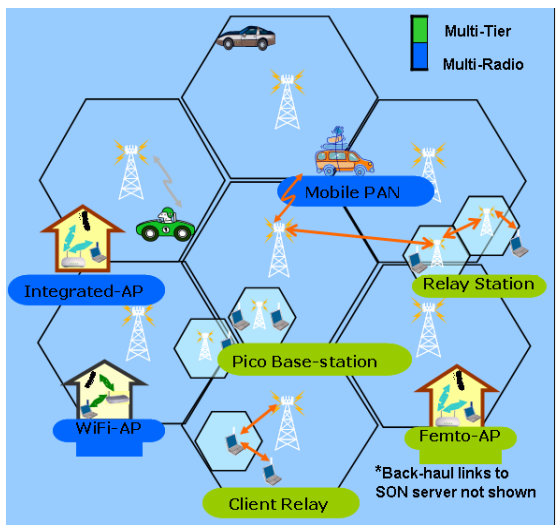

Figure 8.  Hierarchical Structure [29]

1) Macro cell will be marked for fast traffic and micro cell will be marked for slow traffic in hierarchical structure. The RF resources will be dynamically allocated between macro and micro cells on the basis of velocity estimation using adaptive array antennas [24].

2) Pico cell will be marked for hotspots. An adaptive frequency allocation will be applied as strategy to reuse the macro and micro cell resources without frequency planning in indoor picocells [1].

3) Femto cell will be marked as home base stations- which are data access points installed by home users to get better indoor voice and data coverage. Femtocells enable a reduced transmit power, while maintaining good indoor coverage. Penetration losses insulate the femtocell from surrounding femtocell transmissions. As femtocells serve only around 1-4 users, they can devote a larger portion of their resources (transmit power & bandwidth) to each subscriber. A macro / micro cell, on the other hand, has a larger coverage area (500m-20 km radius), and a larger number of users; providing Quality of Service (QoS) for data users is more difficult. Deploying femtocell will enable more efficient uses of precious power and frequency resources [13].

The above proposed structure will undoubtedly enhance the spectral density with the help of diversity and adaptive approach through rake receiver and adaptive antenna respectively. The induction of pico and femto cell will reuse the RF resources of overlaid macro / micro structures which will enhance spectral density manifold. The simple technologies suggested by William C Y Lee for deployment along city streets, deployment along binding roads, deployment under the ground (subway coverage) and in-building designs etc. will also be considered in proposed hierarchical structure [28].

## CONCLUSION

By means of adaptive frequency allocation using adaptive antenna patterns smart / adaptive antennas allow steering the transmit / receive power into certain directions by suppressing the desired power, i.e interference. Additionally the sectorization scheme offers various benefits being combined with SDMA. This has been proved beyond doubt that spectral density will be enhanced substantially. In addition to it, overlaid structures will reduce the handoff using simple handoff algorithms. Moreover RF resources of macro / micro will be reused in pico cell without frequency planning. This will further enhance the capacity of the system beyond any doubt. The use of femto cell will further increase the capacity of the system. In essence proposed network will be a highly efficient cost-effective way of enhancing the capacity and QoS of cellular systems. The above proposed system will be highly effective energy saving cellular network which will reduce power consumption and RF pollution significantly.

AUTHORS PROFILE

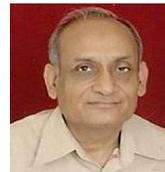

Prof. R. K. Jain has received his BE and ME in 1972 & 1974 from IIT Roorkie, respectively. He has 32 years industrial experience from M/S ITI Ltd. (Top Management) and more than 5 years teaching experience as professor. He is currently working as Prof. and Head of Electronics & Communication Engineering at HIET, Ghaziabad, UP, India. He is currently pursuing PhD degree course from Singhania University as research scholar. His area of research is cellular communication. He has published 9 papers in National / International Conferences and journals.

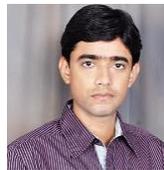

Sumit Katiyar has received his B.Sc. and M.Sc. in 2001 & 2004 respectively from CSJMU, Kanpur, UP, India. He has received his M.Tech. from JIITU, Noida, UP, India. He is currently pursuing PhD degree course from Singhania University as research scholar. He is currently working as Asst. Prof. in department of Electronics & Communication Engineering at HIET, Ghaziabad, UP, India.

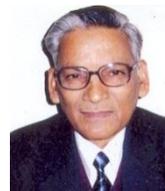

Prof. N. K. Agrawal has received M.Sc. (Electronic) degree from Agra University, Agra, M.Sc (Tech) degree from Birla Institute of Technology, Pilani, Rajsthan and Ph.D degree in Microwave and Radar from Indian Institute of Technology, Roorkee in 1978. He is a senior member of IEEE (USA) and life member of ISTE and ISCEE. He was awarded Young Scientist research award by exchange program between India and France. He has visited many countries. He has worked as visiting professor at University of Shalahddin, Erbil, Iraq. His current Interest includes Antenna, Electromagnetic, Microwave Engineering, and Microwave Absorbers. He has published / presented large number of research papers in these areas.